\documentclass[fleqn,10pt]{wlscirep}
\usepackage[utf8]{inputenc}
\usepackage[T1]{fontenc}

\title{Phonon Thermal Hall Effect: The Roles of Disorder, Annealing, and Metallic Contacts}

\author[1]{Qiaochao Xiang}
\author[1,*]{Xiaokang Li}
\author[1]{Xiaodong Guo}
\author[2,*]{Kamran Behnia}
\author[1,*]{Zengwei Zhu}
\affil[1]{Wuhan National High Magnetic Field Center and School of Physics, Huazhong University of Science and Technology, Wuhan, 430074, China}
\affil[2]{Laboratoire de Physique et d'\'Etude des Mat\'eriaux (ESPCI - CNRS - Sorbonne Universit\'e), PSL Research University, Paris, 75005, France}

\affil[*]{lixiaokang@hust.edu.cn, zengwei.zhu@hust.edu.cn, Kamran.Behnia@espci.fr}

\begin{abstract}
The phonon thermal Hall effect (THE) is a ubiquitous yet poorly understood phenomenon in insulators. Its microscopic origin remains debated, partly due to significant sample-dependent variations that hint at uncontrolled experimental parameters. Using SrTiO$_3$ as a model system, we identify disorder and uncontrolled strain as suppressors of a thermal Hall signal. Crystals with high thermal conductivity exhibit a substantial thermal Hall angle $\nabla T_y / \nabla T_x$ (up to 0.3\% at 9 T), whereas the effect is virtually absent in disordered samples. Crucially, annealing (in air atmosphere) these disordered samples partially restores the THE (approximately 0.1\% at 9 T) with little effect on the longitudinal thermal conductivity. This decoupling reveals that the amplitude of THE is not simply set by the phonon  mean free path. Furthermore, measurements performed  with metallic and insulating contacts yield identical results on the same sample. This definitively rules out parasitic signals as the effect's origin. Our work, by  establishing the phonon THE as an intrinsic property of the crystal lattice and extremely sensitive to disorder,  sharply constrains theoretical scenarios.
\end{abstract}
\begin{document}

\flushbottom
\maketitle
%
%
\thispagestyle{empty}


\section*{Introduction}

The thermal Hall effect (THE)—a transverse heat current in response to a longitudinal thermal gradient under a perpendicular magnetic field—has long been a signature of charge carriers. This paradigm has been challenged by the observation of THE in insulators~\cite{Strohm2005,Onose2010, Hirsch2015,Watanabe2016,Sugii2017, Kasahara2018,Grissonnanche2019,Yamashita2020, Li2020,Grissonnanche2020,Boulanger2020,Sim2021,Chen2022,Uehara2022,Li2023,Chen2024,Chen2024-2,Ataei2024,sharma2024phonon,Meng2024,Jin2025,Li2025}, including non-magnetic~\cite{Li2020,Li2023,sharma2024phonon, Li2025} and elemental ones~\cite{Li2023,Jin2025,Li2025}, implicating phonons as the carriers of this transverse heat. The prevalence of the phonon THE contrasts sharply with the elusive nature of its origin~\cite{Sheng2006, Kagan2008, Zhang2010, Qin2012,Agarwalla2011,Chen2020,Yang2020,Sun2022,Perkins2022,Flebus2022,Mangeolle2022,Guo2022, Behnia2025}. As charge-neutral quasiparticles, phonons are unaffected by the conventional Lorentz force. Theoretical proposals invoke quantum-mechanical mechanisms such as phonon Berry curvature~\cite{Zhang2010, Qin2012}, the intrinsic magnetic moment of chiral phonons~\cite{Zhang2015}, the lattice Aharonov-Bohm effect~\cite{Behnia2025}, the phonon Hall viscosity~\cite{Shragai2025}, and magnetic-field-induced phonon-phonon scattering anisotropy~\cite{Sun2022}. However, the lack of conclusive experimental evidence linking these mechanisms to material properties sustains intense debate.

This theoretical ambiguity is compounded by an experimental crisis of reproducibility. The reported magnitude of the phonon THE varies significantly between studies and even between samples of the same material~\cite{Grissonnanche2019,Grissonnanche2020,Hu2025, Jiang2025}, suggesting a sensitivity to poorly controlled parameters and leading to doubts on the intrinsic nature of the observed signal. Furthermore, it has been proposed that parasitic effects from metallic contacts could be a spurious source~\cite{Ma2025}. Therefore, it is necessary to identify the factors that control the experimental observation of the phonon THE and to answer the following questions: Is it an intrinsic property  or an extrinsic defect-induced phenomenon? How does the amplitude of the signal evolve as a function of disorder, strain and the nature of contacts?

In this work, we address these questions through a systematic study of SrTiO$_3$, the first non-magnetic insulator where the phonon THE was observed~\cite{Li2020}. By tuning sample quality (disorder), thermal annealing (strain), and contact geometry, we establish intrinsic nature of the effect and pinpoint the key factors suppressing it. First, we report a striking sample dependence: high-thermal-conductivity (high-quality) crystals exhibit a substantial thermal Hall angle $\nabla T_y / \nabla T_x$ (up to 0.3\% at 9 T), while the signal is virtually absent in low-thermal-conductivity (more disordered) samples. This suggests that the phonon THE is an intrinsic property of an ideal crystal and that disorder is a key factor suppressing the signal. Secondly, we find that annealing in an air atmosphere can partially restore the phonon THE signal in disordered samples without a concomitant recovery of the longitudinal thermal conductivity. This indicates that the amplitude of THE is not simply set by the phonon mean free path or the concentration of oxygen vacancies, and that the internal strain may be another important suppressor of the signal. Finally, we  resolve the contact controversy by performing simultaneous measurements with both metallic and insulating contacts, obtaining nearly identical results that definitively rule out parasitic signals as the origin.

\section*{Results and Discussions}

\subsection*{Sample Dependence: The Critical Role of Disorder}
\begin{figure}[ht]
\centering
\includegraphics[width=\linewidth]{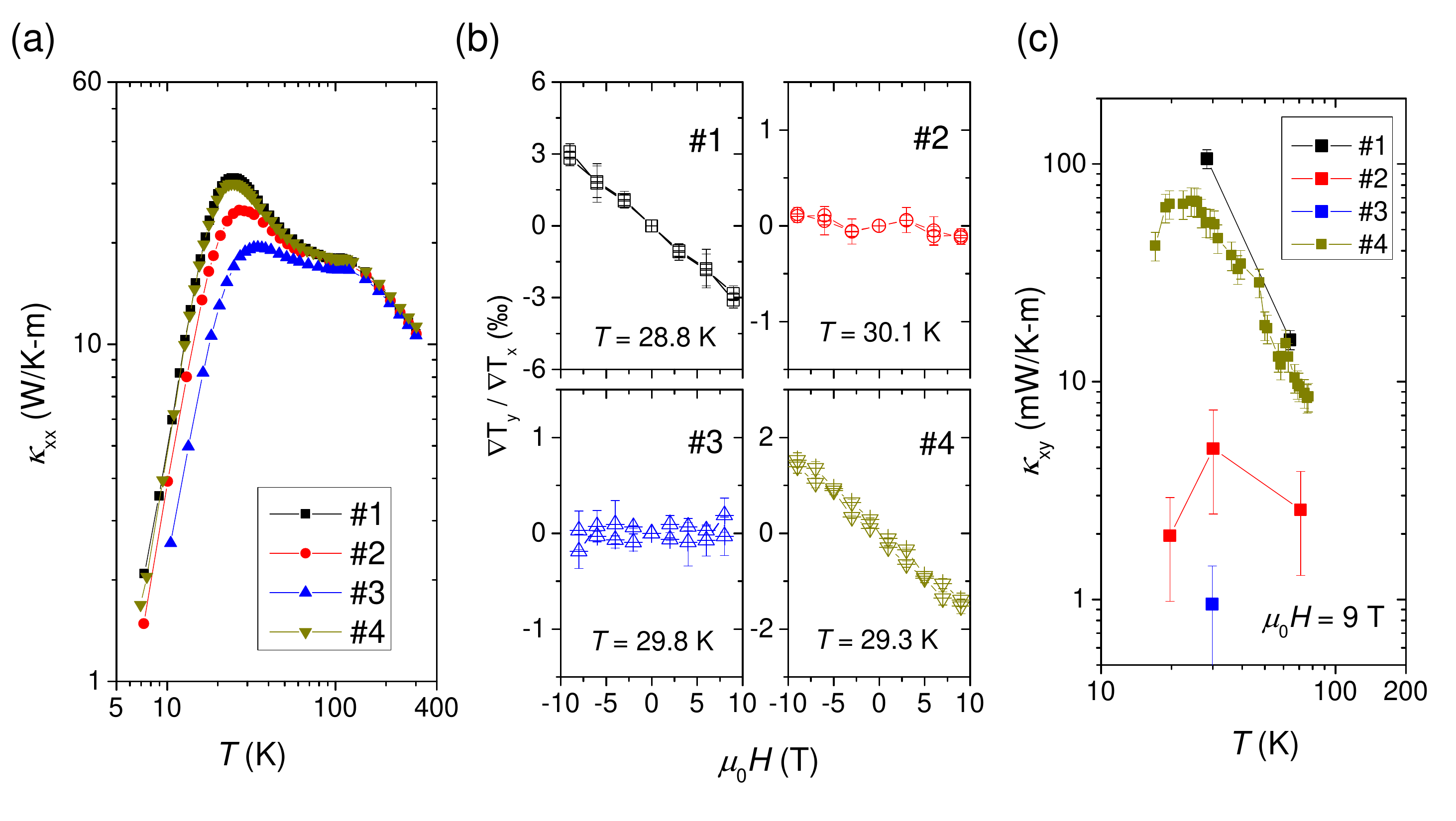}
\caption{\textbf{Sample dependence of thermal transport in SrTiO$_3$.} (\textbf{a}) Temperature-dependent longitudinal thermal conductivity ($\kappa_{xx}$) for samples \#1--\#4. While the curves converge at 300 K, they diverge at lower temperatures, indicating varying levels of disorder. (\textbf{b}) Field dependence of the thermal Hall angle ($\nabla T_y / \nabla T_x$) at $T \approx 29$ K. A strong contrast is observed between the substantial signals in high-quality samples (\#1, \#4) and the virtually absent signals in disordered ones (\#2, \#3). (\textbf{c}) This distinction is further emphasized by the temperature-dependent thermal Hall conductivity ($\kappa_{xy} (T)$), where the response of samples \#1 and \#4 is nearly two orders of magnitude larger.}
\label{fig:sample-dependence}
\end{figure}

To investigate the sample dependence of thermal transport in SrTiO$_3$, we characterized four single crystals (\#1–\#4) sourced from different commercial suppliers. The level of intrinsic disorder—a key factor limiting the phonon mean free path—is known to vary among SrTiO$_3$ sources~\cite{Coey2016, Martelli2018, Trinh2025}. We probed the disorder level directly via the longitudinal thermal conductivity, $\kappa_{xx}$. As shown in Figure~\ref{fig:sample-dependence}a, while the $\kappa_{xx}(T)$ curves converge at room temperature, they diverge significantly at lower temperatures. All samples exhibit a phonon peak near 25 K, consistent with prior studies~\cite{Li2020, Jiang2022, Jin2025}. Samples \#1  and \#4 show a sharp, high $\kappa_{xx}$ peak, indicative of high crystal quality and a long phonon mean free path. In contrast, samples \#2 and \#3 exhibit progressively lower and broader peaks, signaling enhanced phonon scattering due to defects and disorder.

This sample dependence is even more striking in the transverse response. Figure~\ref{fig:sample-dependence}b compares the field-dependent thermal Hall angle ($\nabla T_y / \nabla T_x$), revealing a dramatic disparity: high-quality samples (\#1 and \#4) exhibit substantial signals (up to 0.3\% and 0.15\% at 9 T, respectively), whereas the effect is virtually absent in the more disordered samples (\#2 and \#3). The contrast can be seen more clearly in the temperature-dependent thermal Hall conductivity $\kappa_{xy}(T)$ (Fig.~\ref{fig:sample-dependence}c), where the values for \#1 and \#4 are nearly two orders of magnitude larger than those in \#2 and \#3.  This stark contrast demonstrates that disorder can strongly suppress the phonon thermal Hall effect in SrTiO$_3$. Conversely, this suggests that this effect should be an inherent property of ideal lattices.

The above sample-related results remind us of the experimental discrepancies reported in recent literature~\cite{Grissonnanche2019,Grissonnanche2020,Hu2025, Jiang2025}, and also provide us with a key guiding principle for measuring the thermal Hall effect in insulators: when detecting extreme signal amplitudes—whether very large or very small—we should first assess the quality of the sample crystal, as this can be a significant factor affecting the signal amplitude. 

A natural question then arises: what is the microscopic mechanism by which disorder suppresses the phonon THE? While a detailed investigation is beyond the scope of this work, two important ingredients deserve our attention. The first one is about the internal strain, generated by defects and impurities which inevitably induce lattice distortion~\cite{Meng2024b}. We can speculate that beyond a certain threshold of disorder—evident in our data when the $\kappa_{xx}$ peak falls below 25 W/(K·m) —the resulting spatially varying strain fields may form a mosaic of domains with random orientations. This could lead to a cancellation of the net transverse phonon response, analogous to how magnetic domains cancel out the net magnetization in a ferromagnet. Intriguingly, supporting evidence for such a cancellation mechanism comes from our recent study of the zigzag antiferromagnet NiPS$_3$, where structural twinning was found to cancel out the overall phonon THE~\cite{Meng2024}. The other one is about the quantum paraelectricity. We know that strain~\cite{Haeni2004}, $^{18}$O and Ca substitution all stabilize ferroelectricity. It is known that the latter two roads to ferroelectricity destroy THE~\cite{Sim2021,Jiang2022}.

\subsection*{Annealing Dependence: Decoupling the THE from Thermal Conductivity}
Annealing is an effective method to improve the internal strain condition of crystals. However, vacuum annealing introduces oxygen vacancies and dilute charge carriers into SrTiO$_3$, which can contribute to a phonon-drag thermal Hall signal~\cite{Jiang2022}.  To isolate the influence of internal strain, these extrinsic effects must be avoided. We therefore annealed the disordered samples in air at 1000 $^{\circ}$C~\cite{nabokin2003floating,hanzig2011single}. 

\begin{figure}[ht]
\centering
\includegraphics[width=\linewidth]{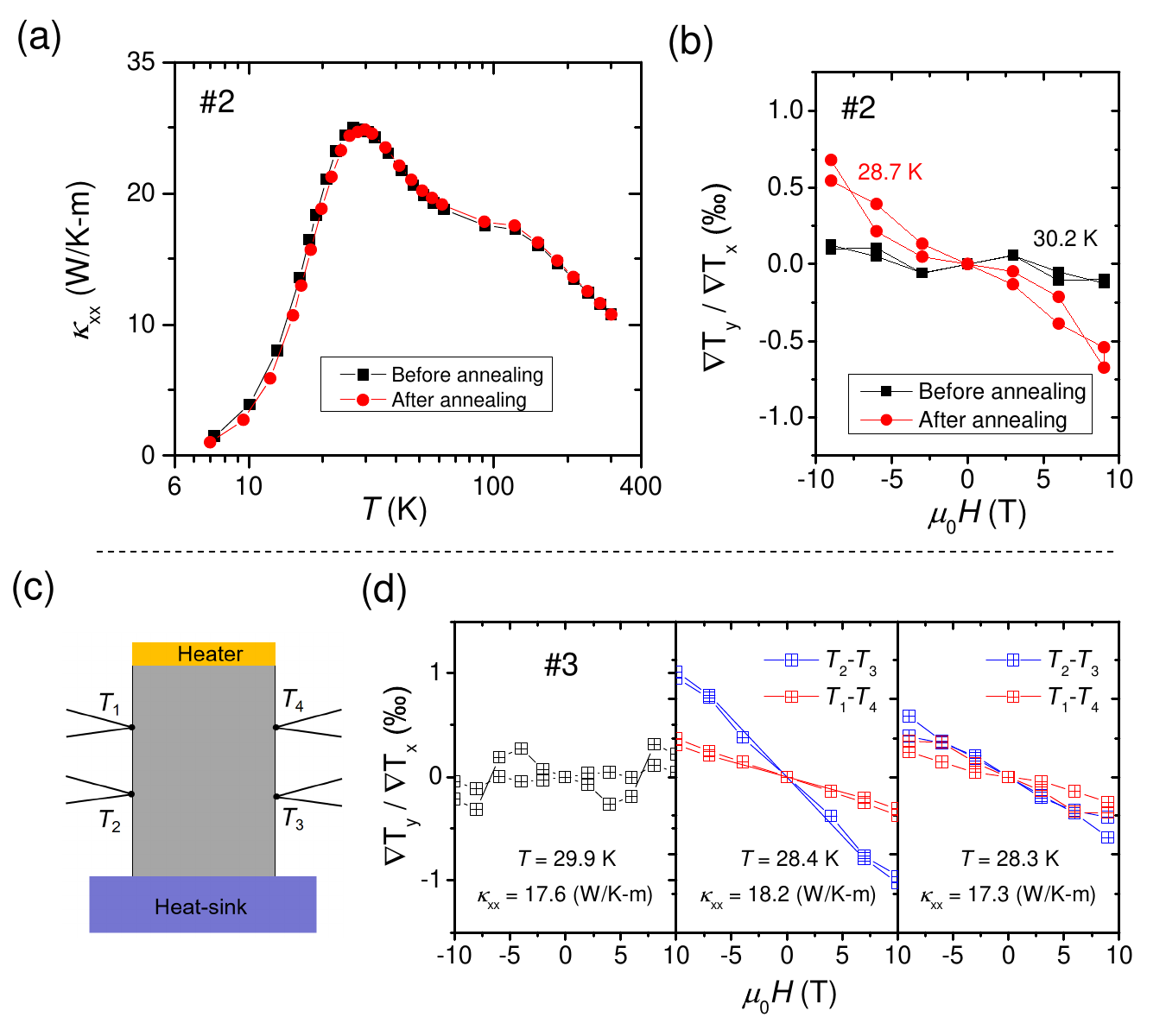}
\caption{\textbf{Recovery of the thermal Hall effect by annealing in disordered SrTiO$_3$.} (\textbf{a}) Temperature dependent longitudinal thermal conductivity ($\kappa_{xx}(T)$) of sample \#2 before and after annealing, showing no significant recovery. (\textbf{b}) In stark contrast, the field-dependent thermal Hall angle ($\nabla T_y / \nabla T_x$), negligible in the unannealed state, emerges clearly after annealing. (\textbf{c}) Schematic of the four-thermocouple measurement geometry. (\textbf{d}) Progressive restoration of the thermal Hall angle ($\nabla T_y / \nabla T_x$) in Sample \#3 with successive annealing cycles: the signal is absent in the unannealed state, partially restored but inhomogeneous after the first short annealing, homogeneous after the second long annealing.}
\label{fig:Annealing}
\end{figure}

Figures~\ref{fig:Annealing}a–b illustrate the effect of annealing on thermal transport in disordered Sample \#2, which was annealed for 24 hours. As shown in Figure~\ref{fig:Annealing}a, the longitudinal thermal conductivity $\kappa_{xx}(T)$ exhibits no significant change after annealing—a trend also observed in Sample \#3 (Figure~\ref{fig:Annealing}d). This indicates that the phonon mean free path, limited by point defects and impurity scattering, remains largely unaffected. Strikingly, however, an obvious thermal Hall angle emerges after annealing (Figure~\ref{fig:Annealing}b), signaling a partial recovery of the phonon thermal Hall effect.

A similar restoration was observed in disordered Sample \#3, which underwent two annealing cycles: a short one (100 minutes at 1000 $^{\circ}$C) followed by a long one (24 hours at 1000 $^{\circ}$C).  The sample was configured with two sets of transverse contacts (T$_1$-T$_4$ and T$_2$-T$_3$), as shown in Figure~\ref{fig:Annealing}c. After the first, short annealing, the THE signal was partially restored but exhibited inhomogeneity between the contact pairs. The subsequent long annealing cycle eliminated this inhomogeneity and stabilized the signal (Figure~\ref{fig:Annealing}d).

These results not only confirm the strong influence of internal strain on the phonon THE, but also reveal a crucial decoupling: annealing restores the transverse thermal response without enhancing the longitudinal thermal conductivity. This implies that internal strain suppresses the THE via a mechanism that operates independently of the phonon mean free path, which typically plays a key role in scattering-related extrinsic mechanisms.

\subsection*{Contact Geometry: Ruling Out Parasitic Signals}
\begin{figure}[ht]
\centering
\includegraphics[width=\linewidth]{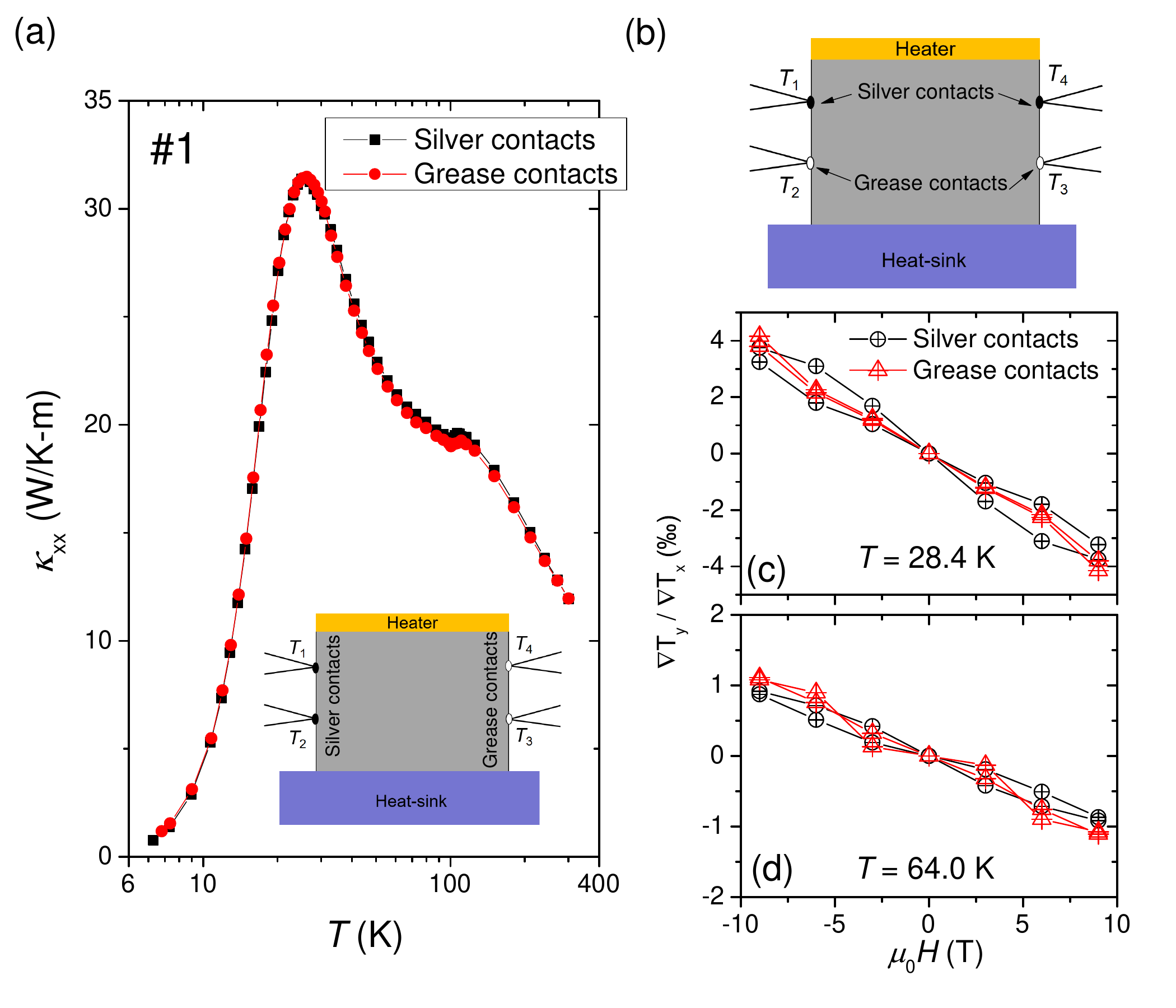}
\caption{\textbf{Validation of contact-independent thermal Hall measurements.} (\textbf{a}) Temperature dependent longitudinal thermal conductivity ($\kappa_{xx}(T)$) measured using metallic (silver paste) and insulating (grease) contacts; the overlapping curves confirm identical longitudinal responses. Inset: contact configuration schematic. (\textbf{b}) Schematic of two pairs of transverse contacts of thermal Hall measurements. (\textbf{c}, \textbf{d}) The field dependent thermal Hall angle ($\nabla T_y / \nabla T_x$) measured at 28.4 K and 64.0 K, respectively. The identical results from both contact types definitively rule out spurious signals originating from the metallic contacts.}
\label{fig:Contacts}
\end{figure}

A recent study has raised the concern that metallic contacts (e.g., silver paste) in thermal Hall measurements may generate substantial spurious signals~\cite{Ma2025}. To address this issue, we performed contact-dependent thermal transport experiments on sample \#1. We first compared the temperature dependent longitudinal thermal conductivity, $\kappa_{xx} (T)$, measured using metallic silver paste contacts on one side and insulating-thermally-conductive grease contacts on the other (see the inset 
 of Figure~\ref{fig:Contacts}a). Comparative measurements revealed nearly identical results (Figure~\ref{fig:Contacts}a), indicating no measurable contact-related artifact in the longitudinal signal. To directly probe the transverse response, we designed a setup for simultaneous thermal Hall measurements using both contact types, as illustrated in Figure~\ref{fig:Contacts}b. Figure~\ref{fig:Contacts}c-d show the field dependent thermal Hall angle ($\nabla T_y / \nabla T_x$) at 28.4 K and 64.0 K respectively. In both temperature points, the results from metallic silver contacts and insulating grease contacts are indistinguishable. This definitively rules out parasitic signals from metallic contacts as the dominant origin of the observed THE, and validates the reliability of the measurement protocol. 
 
 The significant discrepancy between theoretical calculations and experimental measurements of parasitic signals may be due to the model neglecting the influence of interfacial thermal resistance or overestimating the longitudinal and transverse thermal transport coefficients of the silver paste. This interfacial thermal resistance is usually much higher than that of the metallic contacts themselves, while these transport coefficients are usually much lower than those of the silver crystal. However, it should be noted that if the sample's thermal resistance is very large or the measured thermal Hall angle is relatively small—for example, below 0.02\% at 10 T—the influence of the metallic contacts may no longer be negligible. More verification experiments are needed to confirm the signal, such as the setup proposed in the literature for suppressing bypass heat flow~\cite{Ma2025}.

\section*{Conclusion}

In summary, through a systematic investigation of sample quality, annealing, and contact geometry in SrTiO$_3$, we establish that the phonon thermal Hall effect is an intrinsic property of an ideal crystal lattice. Our results demonstrate that its amplitude is decoupled from the phonon mean free path, thereby revealing that disorder and internal strain are potent suppressors of the signal, while also definitively ruling out parasitic effects as its origin. Our work identifies the critical parameters for observing the phonon THE, providing an essential guide for future experiments and placing sharp constraints on theoretical models.

\section*{Samples and Methods}
The SrTiO$_3$ single crystals used in this work were procured from various commercial suppliers. Sample \#1 ($5mm \times 5mm \times 1mm$) was sourced from SurfaceNet. Samples \#2 and \#3 (both $3mm \times 3mm \times 0.5mm$) were obtained from HF-Kejing. Sample \#4, also from HF-Kejing, was fabricated from a crystal grown in Japan. It was cut from a $10mm \times 10mm \times 0.5mm$ mother crystal down to $5mm \times 3mm \times 0.5mm$ for our measurements. 
Samples \#2 and \#3 were annealed in air using a muffle furnace at 1000 $^{\circ}$C for durations ranging from 100 minutes to 24 hours.

All thermal transport experiments were performed in a commercial measurement system (Quantum Design PPMS) within a stable high-vacuum sample chamber. The thermal gradient in the sample was produced through a 4.7 k$\Omega$ chip resistor alimented by a current source (Keithley 6221). The DC voltage on the heater and thermocouples was measured through the DC-nanovoltmeter (Keithley 2182A). Both metallic silver paste and insulating-thermally-conductive grease were used for the contacts.

\section*{Data availability}
The data that support the findings of this study are available from the corresponding author upon reasonable request.


\bibliography{main}



\section*{Acknowledgements}
This work was supported by The National Key Research and Development Program of China (Grant No. 2023YFA1609600, 2024YFA1611200 and 2022YFA1403500), the National Science Foundation of China (Grant No. 12304065, 51821005, 12004123, 51861135104 and  11574097), the Fundamental Research Funds for the Central Universities (Grant No. 2019kfyXMBZ071), the Hubei Provincial Natural Science Foundation ‌(2025AFA072) and the Cai Yuanpei Franco-Chinese program (No. 51258NK).\\

\section*{Author contributions statement}
X.L., Z.Z. and K.B. conceived of and designed the study. Q.X., X.G. performed the thermal transport measurements. X.L., Z.Z., and K.B. analyzed the data. X.L., Z.Z. and K.B. wrote the manuscript with assistance from all the authors.

\section*{Competing interests}
The authors declare no competing interests.

\section*{Additional information}
\textbf{Correspondence} and requests for materials should be addressed to Xiaokang Li, Zengwei Zhu and Kamran Behnia.

\clearpage
\renewcommand{\thesection}{S\arabic{section}}
\renewcommand{\thetable}{S\arabic{table}}
\renewcommand{\thefigure}{S\arabic{figure}}
\renewcommand{\theequation}{S\arabic{equation}}
\setcounter{section}{0}
\setcounter{figure}{0}
\setcounter{table}{0}
\setcounter{equation}{0}

\section*{\centering\LARGE SUPPLEMENTARY NOTES}
\vspace{0.5cm}

\setcounter{figure}{0}

\section{Thermal Hall angle and conductivity}
 Both one-heater-three-thermocouples (type E) and one-heater-four-thermocouples configurations were used in this work. The latter has the advantage of not only allowing simultaneous measurement of both longitudinal ($\nabla T_x = (T_1-T_2)/l$ or $\nabla T_x = (T_4-T_3)/l$) and transverse ($\nabla T_y = (T_2-T_3)/w$ or $\nabla T_y = (T_1-T_4)/w$) thermal gradients, induced by a longitudinal thermal current $J_Q$, but also enables the collection and cross-verification of two sets of transverse temperature difference data ($T_2-T_3$ and $T_1-T_4$) to ensure signal reliability and homogeneity.
 
 The thermal Hall angle is defined as the ratio of transverse to longitudinal thermal gradients ($\nabla T_y / \nabla T_x$). From the measured thermal gradients and Hall angle, the longitudinal ($\kappa_{xx}$) and the transverse ($\kappa_{xy}$) thermal conductivity can be calculated through:
\begin{equation}\label{kappaii}
\kappa_{xx} = \frac{J_Q}{\nabla T_x}
\end{equation}
\begin{equation}\label{kappaij}
\kappa_{xy} = \frac{\nabla T_y}{\nabla T_x} \cdot \kappa_{xx}
\end{equation}
where $l$ denotes the distance between longitudinal thermocouples, $w$ the sample width, and $J_Q$ the heat power per unit cross-sectional area.  The analysis assumes isotropic in-plane thermal conductivity ($\kappa_{xx} = \kappa_{yy}$).

\section{Field-dependent thermal Hall angle in sample \#2}
\begin{figure}[ht]
\centering
\includegraphics[width=\linewidth]{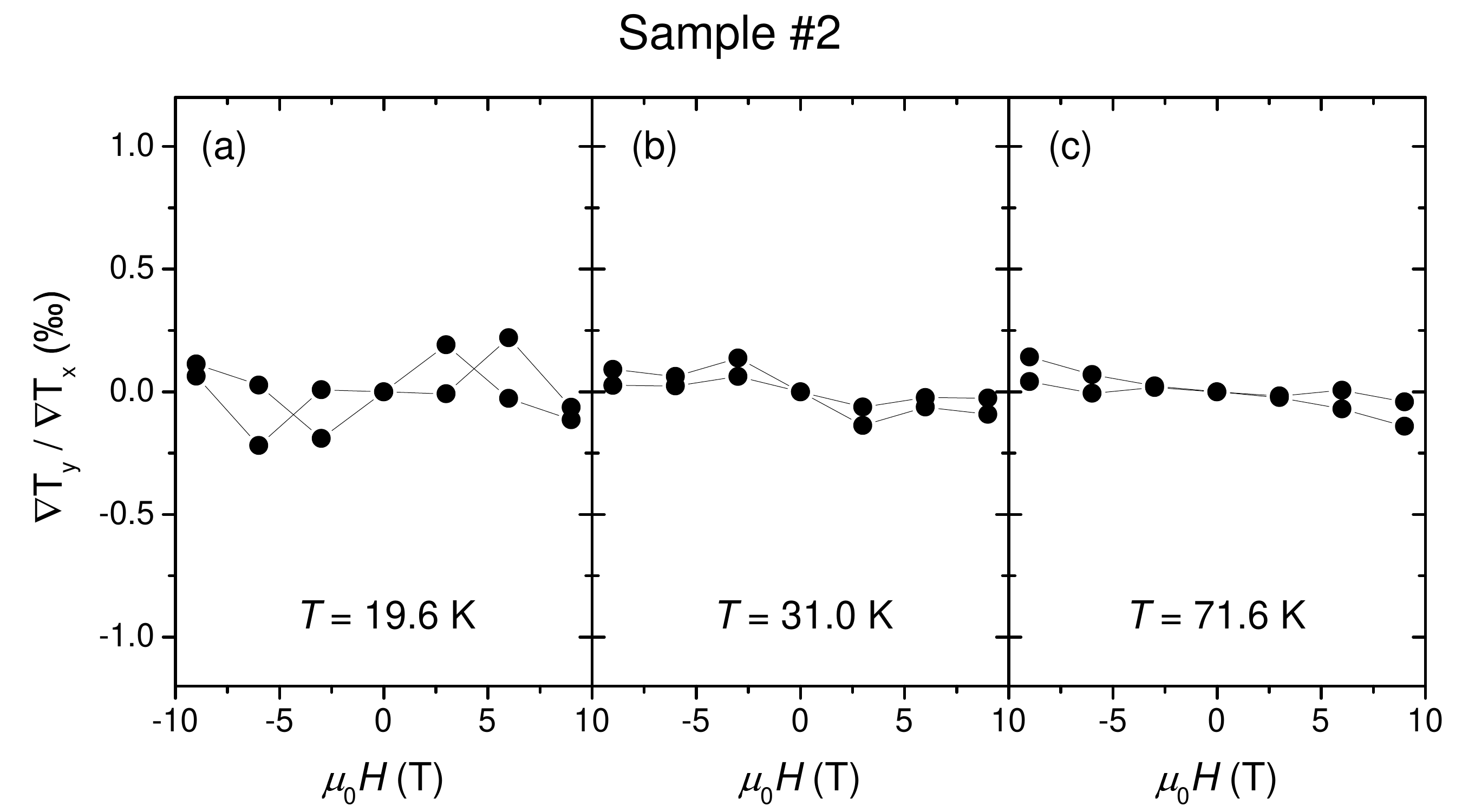}
\caption{\textbf{Thermal Hall angle in sample \#2 before annealing.} (\textbf{a-c}) Field dependence of the thermal Hall angle ($\nabla T_y / \nabla T_x$) measured in Sample \#2 at three different temperatures prior to annealing.}
\label{fig:SM1}
\end{figure}

Figure~\ref{fig:SM1} shows the magnetic field-dependent thermal Hall angle ($\nabla T_y / \nabla T_x$) data for disordered sample \#2 at three temperature points before annealing. All curves show almost undetectable small signals, less than or equal to 0.02\% at 9 T. The noise is relatively high at 19.6 K curve, mainly due to the rapid decrease in coefficient of the type E thermocouple at low temperatures.

\end{document}